\documentclass[twocolumn,english,pra]{revtex4-2}
\usepackage[utf8]{inputenc}
\usepackage{color}
\usepackage{babel}
\usepackage{amsmath}
\usepackage{amsthm}
\usepackage{amssymb}
\usepackage{stackrel}
\usepackage{graphicx}
\usepackage{wasysym}
\usepackage{esint}
\usepackage{xargs}[2008/03/08]
\usepackage[pdfusetitle,
 bookmarks=true,bookmarksnumbered=false,bookmarksopen=false,
 breaklinks=false,pdfborder={0 0 0},pdfborderstyle={},backref=false,colorlinks=true]
 {hyperref}

\makeatletter
\usepackage[justification=raggedright]{caption}
\usepackage{graphicx}

 {
      \theoremstyle{plain}
      }

\usepackage{amsmath}
\usepackage{orcidlink}

\usepackage{ragged2e}
\DeclareCaptionJustification{justified}{\justifying}
\ifdefined\showcaptionsetup
 \PassOptionsToPackage{caption=false}{subfig}
\fi
\usepackage{subfig}
\makeatother

\begin{document}
\title{Achieving Heisenberg scaling by probe-ancilla interaction in quantum
metrology}
\author{Jingyi Fan\orcidlink{0000-0001-5816-1235}$^{1}$}
\author{Shengshi Pang\orcidlink{0000-0002-6351-539X}$^{1,2}$}
\email{pangshsh@mail.sysu.edu.cn}

\affiliation{$^{1}$School of Physics, Sun Yat-sen University, Guangzhou, Guangdong
510275, China}
\affiliation{$^{2}$Hefei National Laboratory, University of Science and Technology
of China, Hefei 230088, China}
\begin{abstract}
The Heisenberg scaling is an ultimate precision limit of parameter
estimation allowed by the principles of quantum mechanics, with no
counterpart in the classical realm, and has been a long-pursued goal
in quantum metrology. It has been known that interactions between
the probes can help reach the Heisenberg scaling without entanglement.
In this paper, we show that interactions between the probes and the
additional dimensions of an ancillary system may also increase the
precision of parameter estimation to surpass the standard quantum
limit and attain the Heisenberg scaling without entanglement, if the
measurement scheme is properly designed. The quantum Fisher information
exhibits periodic patterns over the evolution time, implying the existence
of optimal time points for measurements that can maximize the quantum
Fisher information. By implementing optimizations over the Hamiltonian,
the initial states of the probes and the ancillary system, the interaction
strength, and the time points for measurements, our protocol achieves
the Heisenberg scaling for the parameter of the probe Hamiltonian,
in terms of both evolution time and probe number. Our protocol features
two aspects: (i) the Heisenberg scaling can be achieved by a product
state of the probes and (ii) mere local measurement on the ancilla
is sufficient, both of which reduce the quantum resources and the
implementation complexity to achieve the Heisenberg scaling. The paper
is concluded by the investigation of the effects of noise on this
protocol.
\end{abstract}
\maketitle
\newcommandx\tr[1][usedefault, addprefix=\global, 1=]{\mathrm{Tr}_{ #1 }}%
\newcommandx\abs[1][usedefault, addprefix=\global, 1=]{\left|#1\right|}%
\newcommandx\bra[1][usedefault, addprefix=\global, 1=]{\langle#1\rvert}%
\newcommandx\ket[1][usedefault, addprefix=\global, 1=]{\lvert#1\rangle}%
\newcommandx\braket[2][usedefault, addprefix=\global, 1=, 2=]{\langle#1\rvert#2\rangle}%
\newcommandx\aver[1][usedefault, addprefix=\global, 1=]{\langle#1\rangle}%
\newcommandx\vt[1][usedefault, addprefix=\global, 1=]{\boldsymbol{#1}}%
\global\long\def\wp{\omega}%
\global\long\def\wa{\omega_{a}}%
\global\long\def\hp{H_{p}}%
\global\long\def\ha{H_{a}}%
\global\long\def\rp{\rho_{p}}%
\global\long\def\ra{\rho_{a}}%
\global\long\def\rpa{\rho_{pa}}%
\global\long\def\psp{\psi_{p}}%
\global\long\def\psa{\psi_{a}}%
\newcommandx\php[1][usedefault, addprefix=\global, 1=]{\phi_{p}^{#1}}%

\section{Introduction}

The pursuit of precision in measurements is a long-standing goal in
physics and other fields and has been extensively studied in the science
of metrology. Quantum metrology, the extension of metrology to the
quantum regime, aims to push the measurement precision to the ultimate
limit constrained by the intrinsic statistical uncertainty introduced
by quantum mechanics. A fundamental limit in quantum metrology is
the standard quantum limit on the errors of parameter estimation by
measurements of uncorrelated systems, which scales linearly with $N^{-1/2}$
with $N$ the total number of measurements. The standard quantum limit
is essentially rooted in the central limit theorem of classical statistics
\citep{kay1993fundamentals}. In contrast, if non-classical resources
such as quantum entanglement and quantum squeezing \citep{caves1981quantummechanical,wineland1992spinsqueezing}
are introduced to the systems, the sensitivity of quantum measurements
can be enhanced and the precision limit of parameter estimation may
surpass the standard quantum limit and reach the Heisenberg limit.
The Heisenberg limit is a precision scaling of parameter estimation
inverse to the evolution time or the number of probes, much lower
than the standard quantum limit with no counterpart in classical physics.
This advancement has lead to substantial progress in quantum metrology
\citep{giovannetti2011advances}, both theoretically and experimentally,
as reported in previous literatures concerning typical tasks such
as phase estimation \citep{toth2014quantum,pezz`e2018quantum,gianani2020assessing,lee2006adiabatic}
and more recently focused on multi-parameter estimation \citep{albarelli2020aperspective,demkowicz-dobrzanski2020multiparameter,szczykulska2016multiparameter},
etc. In addition, quantum metrology has been applied in a variety
of areas such as atomic clocks \citep{nichol2022anelementary,ludlow2015optical,katori2011optical},
gravitational wave sensing \citep{caves1981quantummechanical,schnabel2010quantum},
dark matter \citep{malnou2019squeezed,bass2024quantum}, biological
sensing \citep{taylor2016quantum} and quantum imaging \citep{perez-delgado2012fundamental,lugiato2002quantum,albarelli2020aperspective,genovese2016realapplications,moreau2019imaging},
revolutionizing our ability to collect information on physical quantities
and parameters with high precision.

Parameter estimation is a fundamental component of quantum metrology,
as it offers the appropriate tools and methods to develop quantum
measurement schemes that achieve high precision \citep{helstrom1969quantum,braunstein1994statistical,holevo2011probabilistic}.
In classical parameter estimation, the inference of unknown parameters
relies on parameter-dependent probability distributions. The Cramér-Rao
bound sets a lower limit on the variance of parameter estimation over
all estimation strategies, which turns out to be determined by the
inverse of the Fisher information \citep{helstrom1969quantum}. In
the quantum regime, the parameters of interest are usually encoded
into the quantum state of the system. The quantum Fisher information,
defined as the maximum Fisher information over all possible generalized
measurements, gives the quantum Cramér-Rao bound for quantum parameter
estimation. In addition, quantum estimation protocols can take advantage
of quantum mechanical features, such as non-classical properties of
quantum states, which offer enhanced sensing capability compared to
classical methods.

In quantum metrology, the parameters of interest are often associated
with quantum dynamics, and the primary goal is to design measurement
strategies that achieve the highest precision. A general approach
is to encode the parameters of interest into the probe system through
an associated quantum dynamical evolution, and then to measure the
evolved system and estimate the parameters through processing the
measurement outcomes \citep{giovannetti2006quantum,braun2018quantumenhanced}.
To maximize the information of the parameters that can be extracted
from the evolved probe system, it is essential to optimize both the
state of the probe system and the measurement strategies, as they
significantly influence the precision of the estimation \citep{humphreys2013quantum,baumgratz2016quantum}.
In addition to these quantum metrology protocols, the sensitivity
of the precision can also be enhanced by utilizing appropriate interactions
\citep{boixo2007generalized,demkowicz-dobrzanski2014usingentanglement},
introducing quantum controls \citep{yang2022variational,pang2017optimal,liu2017quantum,yuan2015optimal},
etc. With appropriate initial states and measurement strategies, the
precision limit can exceed the standard quantum limit and achieve
the Heisenberg scaling for typical tasks of quantum parameter estimation
such as phase estimation \citep{giovannetti2006quantum}.

In practical scenarios, the environment-induced noise can undermine
the advantage of quantum metrology and prevents measurements from
attaining the Heisenberg limit, even reducing it to the standard quantum
limit \citep{huelga1997improvement,demkowicz-dobrzanski2009quantum,shaji2007qubitmetrology,escher2011general}.
In the presence of noise, the finite coherence time reduces the period
during which the quantum states can maintain the superposition and
the quantum Fisher information can increase, suggesting the importance
of choosing appropriate time points for measurements to extract as
much information as possible from noisy quantum systems, which is
essential for the design of robust quantum parameter estimation strategy
against environmental noises. The pursuit of measurement precision
that exceeds the standard quantum limit has led to the development
of a variety of ingenious strategies, including the use of squeezed
states \citep{caves1981quantummechanical,wineland1992spinsqueezing,ma2009fisherinformation,hyllus2010entanglement,rozema2014optimizing,hosten2016measurement},
ancillary systems \citep{he2021quantum,boixo2008quantumlimited,boixo2007generalized,demkowicz-dobrzanski2014usingentanglement,chen2024qubitassisted,zhang2022approaching,ullah2023lowtemperature},
the exploitation of the time-reversal procedure \citep{agarwal2022quantifying,linnemann2016quantumenhanced,nolan2017optimal,colombo2022timereversalbased},
non-Markovian effects \citep{chin2012quantum,smirne2016ultimate,kukita2021heisenberglimited},
etc.

\begin{figure}
\subfloat[\label{fig:=000020circuit-1}General protocol \centering]{\includegraphics[width=8.8cm]{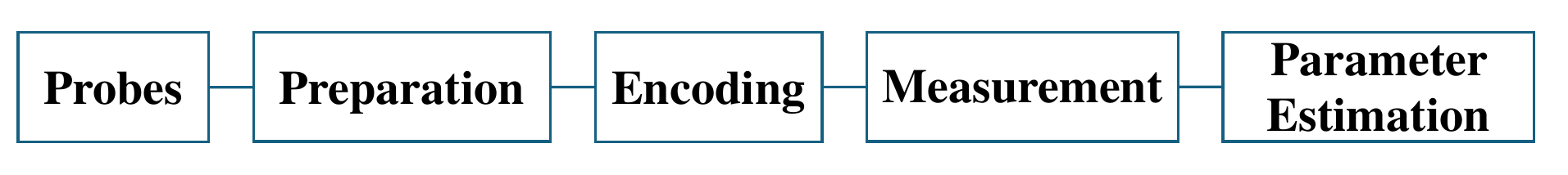}

}

\subfloat[\label{fig:=000020circuit-2} Ancilla-assisted protocol\centering]{\includegraphics[width=8.8cm]{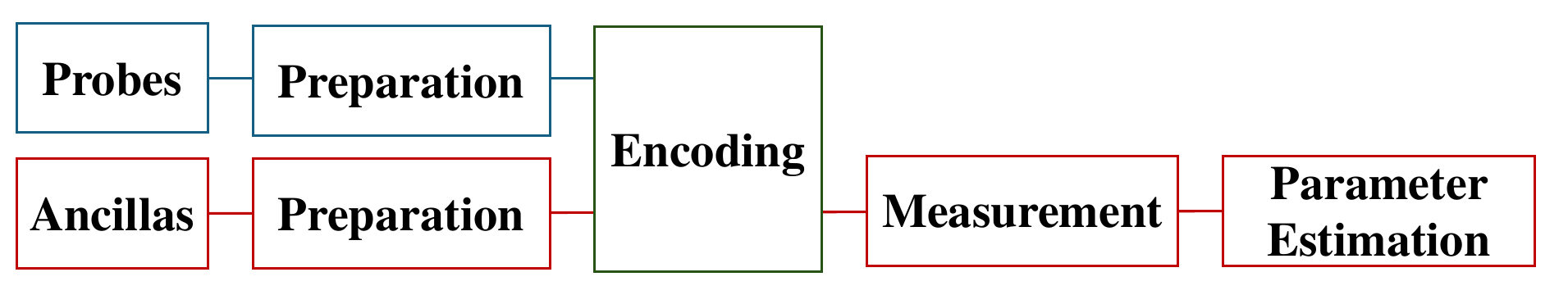}

}

\caption{Schematic diagrams outlining a general quantum metrology protocol
and the current ancilla-assisted protocol, both including preparing
quantum probes in a desired initial state, allowing the probes to
evolve to encode the parameters of interest, and extracting the information
of the parameters from the measurements. (a) The general protocol
takes advantage of quantum resources such as entanglement and squeezed
states to raise the measurement precision beyond the standard quantum
limit approaching the Heisenberg scaling. (b) An alternative protocol
using ancillas and uncorrelated probes employs probe-ancilla interaction
to achieve the Heisenberg scaling without entanglement. This protocol
also simplifies the measurement process by observing the ancillas
only.}
\end{figure}

A general approach of quantum metrology is illustrated in Fig. \ref{fig:=000020circuit-1},
which involves preparing the probes in a proper initial state, encoding
the parameters of interest through a unitary transformation, performing
measurements on the final state, and processing the outcomes by proper
estimation strategies to infer the unknown parameters. Typical quantum
protocols include the utilization of quantum resources like entangled
states to enhance the measurement sensitivity. To date, the multi-particle
entanglement has been realized in various many-body quantum systems
\citep{bouwmeester1999observation,yao2012observation,zhong2021deterministic,monz201114qubit,neumann2008multipartite},
leading to enhanced measurement precision \citep{grote2013firstlongterm,grangier1987squeezedlightenhanced,xie2021beating}.
Nevertheless, the use of quantum entanglement often involves the challenges
regarding the cost, efficiency, and quality in their preparation and
application. As the size of the system scales up, the quantum resources
become increasingly delicate and fragile against environmental noise,
which can disrupt the non-classical characteristics of the quantum
systems even if the noise is weak. Such vulnerability gives rise to
significant decoherence in quantum states, resulting in the loss of
quantum entanglement and the return to classical behavior of quantum
systems \citep{dawson2005entanglement,wang2009entanglement}. 

In recent years, interactions have been found useful in quantum metrology
to achieve or even beat the Heisenberg limit \citep{roy2008exponentially},
without the assistance of quantum entanglement, circumventing the
difficulties of preparing quantum correlated states. It is shown that
$k$-body interactions between the probes can increase the precision
of quantum parameter estimation beyond the Heisenberg scaling and
push the precision limit to a super-Heisenberg scaling $N^{-k}$ \citep{boixo2007generalized,boixo2008quantumlimited,boixo2008quantum}.
Such interaction-based super-Heisenberg scalings have been realized
by experiments \citep{napolitano2011interactionbased,hou2021superheisenberg}.

Inspired by the previous literatures on interaction-based quantum
metrology, in this paper, we propose a simple but non-trivial metrological
protocol that does not require interaction or entanglement between
the probes but introduces an ancillary system to interact with them.
The interaction between the ancillary system and the probe system
is mediated by a two-body coupling Hamiltonian, which can be experimentally
realized in a variety of quantum systems including trapped ions \citep{porras2004effective,lu2019globalentangling,bohnet2016quantum},
superconducting circuits \citep{yan2019strongly,song2019generation,roushan2017spectroscopic,ye2019propagation},
etc. We show that such two-body interaction between the ancillary
system and the probe system can lead to enhanced precision that surpasses
the standard quantum limit in estimating the parameters of the probes,
and the measurement to realize such enhanced precision just needs
to be local on the ancillary system, without the necessity of nonlocal
measurements as in general quantum metrological protocols. Additionally,
the quantum Fisher information of the evolved ancillary system exhibits
periodic patterns over the evolution time. In order to achieve the
highest precision, optimizations are required on the tailoring of
the Hamiltonian, the time points for measurements, and the tuning
of the interaction strength between the ancillary system and the probe
system. By implementing these optimizations, the precision of the
protocol can achieve the Heisenberg scaling at those periodic time
points, in terms of both evolution time and probe number. The whole
protocol is illustrated in Fig. \ref{fig:=000020circuit-2}. As noise
can generally destroy the coherence of quantum systems and undermine
the advantage of quantum technologies, we consider the effects of
noise on this protocol, and analyze the behavior of quantum Fisher
information in a dephasing process as an example.

The paper is organized as follows. In Sec. \ref{sec:Preliminaries},
we give preliminaries for the general theory of parameter estimation
in quantum metrology. In Sec. \ref{sec:Hamiltonian}, we introduce
the general framework for our metrology protocol. Optimizations of
the initial states of both the ancilla and the probe system, the Hamiltonian,
the time points for measurements, and the interaction strength are
investigated, and the highest precision of the protocol is derived
in Sec. \ref{sec:optimization}. In Sec. \ref{sec:noise}, we study
the influence of noise on our protocol and show the long-term behavior
of quantum Fisher information in the presence of a dephasing process.
Finally the paper is concluded in Sec. \ref{sec:Conclusion}.

\section{Preliminaries\protect\label{sec:Preliminaries}}

In this section, we provide a concise overview of the fundamental
concepts in quantum metrology relevant to the current research.

In quantum metrology, the key task is to improve the estimation precision
for unknown parameters in quantum systems. Enhanced sensitivity can
be achieved if typical quantum resources, such as entanglement and
squeezed states, are utilized. The unknown parameters are encoded
into the state of some probes by quantum processes dependent on those
parameters. Subsequently, appropriate measurements are performed on
the probes, and the measurement outcomes are analyzed by sophisticated
statistical strategies to infer the unknown parameters.

The theory of quantum parameter estimation is rooted in the classical
theory of parameter estimation. In the classical parameter estimation,
the goal is to estimate an unknown parameter $\theta$ from a $\theta$-dependent
probability distribution $p\left(x|\theta\right)$ of a random variable
$x$. The mean squared error of the estimator $\hat{\theta}$ is limited
by the Cramér--Rao bound\textcolor{blue}{{} \citep{cramer1946mathematical}},
\begin{equation}
{\rm E}[(\hat{\theta}-\theta)^{2}]\geq\left(NF\right)^{-1}+{\rm E}[(\hat{\theta}-\theta)]^{2},
\end{equation}
where the first term at the right side is the lower bound of the variance
of the estimation and the second term is the systematic error of the
estimation, $N$ is the number of measurement repetitions, and $F$
represents the Fisher information, defined as
\begin{equation}
F=\int\frac{\left[\partial_{\theta}p\left(x|\theta\right)\right]^{2}}{p\left(x|\theta\right)}dx.
\end{equation}

When the parameter estimation theory is applied to the quantum realm,
the parameter of interest is usually encoded in a quantum state, and
the choice of measurement basis can significantly influence the distribution
of measurement outcomes and change the precision of parameter estimation.
The quantum Cramér-Rao bound, derived by minimizing the variance of
estimation over all possible measurement bases, gives the lowest achievable
variance for any estimation strategy. This extends the classical Fisher
information to the quantum Fisher information which quantifies the
sensitivity of parameter estimation in quantum systems.

For an unknown parameter $\theta$ in a density matrix $\rho_{\theta}$,
the quantum Fisher information can be derived as
\begin{equation}
F_{Q}=\tr\left(\rho_{\theta}L_{\theta}^{2}\right),
\end{equation}
where the Hermitian operator $L_{\theta}$ is the symmetric logarithmic
derivative with respect to the parameter $\theta$ \citep{braunstein1994statistical},
determined by
\begin{equation}
\partial_{\theta}\rho_{\theta}=\frac{1}{2}\left(\rho_{\theta}L_{\theta}+L_{\theta}\rho_{\theta}\right).
\end{equation}
Consider an initial state with a spectral decomposition $\rho=\sum_{i}p_{i}\ket[\lambda_{i}]\bra[\lambda_{i}]$,
the quantum Fisher information can be explicitly derived as
\begin{equation}
F_{Q}=\sum_{\lambda_{i}+\lambda_{j}\neq0}\frac{2\abs[{\bra[\lambda_{i}]\partial_{\theta}\rho_{\theta}\ket[\lambda_{j}]}]^{2}}{\lambda_{i}+\lambda_{j}}.\label{eq:=000020FQ}
\end{equation}

For a pure state $\rho_{\theta}=\ket[\psi_{\theta}]\bra[\psi_{\theta}]$,
the quantum Fisher information \eqref{eq:=000020FQ} is reduced to
\begin{equation}
F_{Q}=4\left[\braket[\partial_{\theta}\psi_{\theta}][\partial_{\theta}\psi_{\theta}]-\abs[{\braket[\psi_{\theta}][\partial_{\theta}\psi_{\theta}]}]^{2}\right],\label{eq:=000020FQ-pure}
\end{equation}
with $\ket[\partial_{\theta}\psi_{\theta}]=\partial\ket[\psi_{\theta}]/\partial\theta$.
Furthermore, for a quantum system evolved from a pure state $\ket[\psi_{0}]$
under a $\theta$-dependent Hamiltonian $H_{\theta}$ for time $t$,
the quantum Fisher information can be further simplified to
\begin{equation}
F_{Q}=4t^{2}\aver[\Delta^{2}H_{\theta}],\label{eq:=000020FQ-pure-unitary}
\end{equation}
where $\aver[\Delta^{2}H_{\theta}]$ is the variance of $H_{\theta}$
for the initial state $\ket[\psi_{0}]$, given as
\begin{equation}
\aver[\Delta^{2}H_{\theta}]=\bra[\psi_{0}]H_{\theta}^{2}\ket[\psi_{0}]-\bra[\psi_{0}]H_{\theta}\ket[\psi_{0}]^{2}.
\end{equation}

A simple and typical scenario of quantum metrology considers a probe
with a Hamiltonian $H_{\theta}=\theta\widetilde{H}$. The maximal
quantum Fisher information for the parameter $\theta$ can be obtained
as
\begin{equation}
\max_{\ket[\psi_{0}]}F_{Q}=t^{2}\left(\lambda_{\text{max}}-\lambda_{\text{min}}\right)^{2},
\end{equation}
when
\begin{equation}
\ket[\psi_{0}]=\left(\ket[\lambda_{\text{max}}]+\ket[\lambda_{\text{min}}]\right)/\sqrt{2},
\end{equation}
where $\lambda_{\text{max}}$ and $\lambda_{\text{min}}$ are the
maximum and minimum eigenvalues of the Hamiltonian $\widetilde{H}$,
corresponding to eigenstates $\ket[\lambda_{\text{max}}]$ and $\ket[\lambda_{\text{min}}]$.

For a system of $N$ probes, if the probes are uncorrelated, the maximal
quantum Fisher information is
\begin{equation}
\max_{\ket[\psi_{0}]}F_{Q}=Nt^{2}\left(\lambda_{\text{max}}^{(i)}-\lambda_{\text{min}}^{(i)}\right)^{2},\label{eq:=000020QFI-N}
\end{equation}
with the initial state
\begin{equation}
\ket[\psi_{0}]=\left(\ket[\lambda_{\text{max}}]+\ket[\lambda_{\text{min}}]\right)^{\otimes N}/\sqrt{2},
\end{equation}
according to the additivity of quantum Fisher information for uncorrelated
systems \citep{pezz`e2018quantum}, implying the measurement precision
scales as $\Delta\theta\wasypropto1/\sqrt{N}$, which is the standard
quantum limit. However, by employing a maximally entangled state
\begin{equation}
\ket[\psi_{0}]=\left(\ket[\lambda_{\text{max}}]^{\otimes N}+\ket[\lambda_{\text{min}}]^{\otimes N}\right)/\sqrt{2},
\end{equation}
the maximal quantum Fisher information becomes
\begin{equation}
\max_{\ket[\psi_{0}]}F_{Q}=N^{2}t^{2}\left(\lambda_{\text{max}}^{(i)}-\lambda_{\text{min}}^{(i)}\right)^{2},\label{eq:=000020QFI-N2}
\end{equation}
indicating that the standard quantum limit can be surpassed \citep{dariano2001usingentanglement}
and higher precision is attained by the measurements, and the corresponding
measurement precision scales as $\Delta\theta\wasypropto1/N$, which
is the Heisenberg scaling, a $\sqrt{N}$ improvement over the standard
quantum limit.

The simplest scenario involves a two-level system, for which the Bloch
representation is employed to depict the quantum state, which can
be represented by $\rho_{\theta}=\frac{1}{2}\left(I+\vt[r]_{\theta}\cdot\vt[\sigma]\right)$,
where $\boldsymbol{r}_{\theta}=[x_{\theta},y_{\theta},z_{\theta}]$
is the Bloch vector satisfying $\abs[\text{\ensuremath{\vt[r]}}_{\theta}]^{2}\leq1$,
and $\vt[\sigma]=[\sigma_{x},\sigma_{y},\sigma_{z}]$ is the vector
of the Pauli matrices. For a two-level system, the quantum Fisher
information can then be written as
\begin{equation}
F_{Q}=\abs[\partial_{\theta}\text{\ensuremath{\vt[r]}}_{\theta}]^{2}+\frac{\left(\text{\ensuremath{\vt[r]}}_{\theta}\cdot\partial_{\theta}\text{\ensuremath{\vt[r]}}_{\theta}\right)^{2}}{1-\left|\text{\ensuremath{\vt[r]}}_{\theta}\right|^{2}},\label{eq:=000020FQ=000020bloch}
\end{equation}
where $\left|\cdot\right|$ denotes the Euclidean norm of a vector.
The Bloch representation, compared with Eq.\eqref{eq:=000020FQ},
is a convenient tool as it obviates the need for spectral decomposition
of the density matrix, a process that can be resource-consuming and
hard to tackle for high-dimensional systems.

\section{Hamiltonians and Dynamics of ancillary system\protect\label{sec:Hamiltonian}}

In this paper, we consider a scenario where an ancillary system is
introduced to interact with the probes and investigate its effect
in enhancing the sensitivity of measurement. The general Hamiltonian
for the probe system and the ancillary system involving the free Hamiltonian
$H_{0}$ and interaction Hamiltonian $H_{I}$ can be written as
\begin{equation}
H_{\text{tot}}=H_{0}+H_{I}=\wp\hp+\wa\ha+gB\otimes A,\label{eq:htot}
\end{equation}
where $\hp$ and $\ha$ are the free Hamiltonians of the probe system
and the ancillary system,\textcolor{red}{{} }$\wa$ is the frequency
of the ancillary system, and $\wp$ is the frequency of the probe
system which is supposed to be the unknown parameter to estimate.
The interaction Hamiltonian $H_{I}$ describes the coupling between
the two systems with the strength $g$, where $B$ and $A$ are the
observables of the probe system and the ancillary system respectively.

The goal is to measure the parameter $\wp$ of the Hamiltonian $\hp$
in the probe system. Suppose the initial density matrix can be factorized
between the probe system and ancillary system, $\rpa\left(0\right)=\rp\left(0\right)\otimes\ra\left(0\right)$.
The joint evolution of the probe system and the ancillary system can
be written as
\begin{equation}
\rpa\left(t\right)=U\left(t\right)\rp\left(0\right)\otimes\ra\left(0\right)U^{\dagger}\left(t\right),\label{eq:=000020rho-sa}
\end{equation}
where $U\left(t\right)=e^{-iH_{\text{tot}}t}$ is the unitary dynamical
evolution under the total Hamiltonian \eqref{eq:htot} encoding the
parameter $\wp$ into the evolved state. 

As we focus on local measurements on the ancillary system alone in
this paper, the probe system can be traced out from the joint density
matrix \eqref{eq:=000020rho-sa}, and the reduced evolution of the
ancillary system can be written as
\begin{equation}
\ra\left(t\right)=\mathrm{Tr}_{p}\left[\rpa\left(t\right)\right].
\end{equation}
As the reduced density matrix $\ra\left(t\right)$ is dependent on
the unknown parameter $\omega$, the information of the parameter
$\wp$ can be obtained through the measurements on the ancillary system.
The precision of estimating the parameter $\omega$ can be determined
by the quantum Fisher information for a general mixed state \eqref{eq:=000020FQ}.
When the ancilla is a two-dimensional system, the quantum Fisher information
can be simplified in the Bloch representation \eqref{eq:=000020FQ=000020bloch}.

\section{Optimizations of quantum Fisher information\protect\label{sec:optimization}}

In quantum metrology, the parameter of interest is inferred from the
evolved state of the system and the evolved state of the system generally
depends on the initial state of the system and the evolution path
under the Hamiltonian, so proper optimizations of the initial state
and the Hamiltonian are vital to the performance of measurements precision.

In this section, we investigate the optimizations of the estimation
precision over the initial state and the Hamiltonian of the probes
and the ancillary system, and show that such optimizations can significantly
increase the estimation precision. In particular, when there are multiple
probes present for sensing the frequency, proper optimizations of
the interaction Hamiltonian between the probes and the ancillary system
can raise the precision limit to the Heisenberg scaling even when
the probes are initially in a product state, which manifests the power
of interaction between the probes and the external degrees of freedom
introduced by the ancilla, in addition to the known advantage of interaction
within the probes \citep{boixo2007generalized,boixo2008quantum,boixo2008quantumlimited,roy2008exponentially,napolitano2011interactionbased}.

\subsection{Optimization of ancilla state}

A proper initial state is crucial to enhancing the precision of parameter
estimation for the quantum system. While the probe system is generally
considered to have arbitrary dimensions, only the highest and the
lowest levels of the system will be involved in determining the maximal
Fisher information \citep{giovannetti2006quantum}. Hence, we use
two-level systems to simulate the probes in this paper. For simplicity,
we also assume the ancillary system to be two-level, which is not
necessarily the optimal choice but will be shown to be sufficient
to attain the Heisenberg scaling. For the convenience of computation,
we choose the observable $A$ of the ancillary system in the interaction
Hamiltonian to commute with the free Hamiltonian $\ha$ of the ancillary
system, i.e., $\left[\ha,A\right]=0$. This allows $\ha$ and $A$
to share a set of common eigenstates $\text{\ensuremath{\ket[a_{k}]}}$,
$k=1,2$, associated with eigenvalues $h_{k}$ and $a_{k}$, respectively.

Suppose the ancillary system is initialized in a pure state: 
\begin{equation}
\ket[\psi_{a}]=\cos\alpha\ket[a_{1}]+e^{-i\phi}\sin\alpha\ket[a_{2}].
\end{equation}
Its reduced density matrix after an evolution under the total Hamiltonian
\eqref{eq:htot} for time $t$ is given by
\begin{equation}
\ra\left(t\right)=\mathrm{Tr}_{p}\left[e^{-iH_{\text{tot}}t}\rp\left(0\right)\otimes\ket[\psa]\bra[\psa]e^{iH_{\text{tot}}t}\right].
\end{equation}
Since the eigenstates $\text{\ensuremath{\ket[a_{1}]}}$and $\text{\ensuremath{\ket[a_{2}]}}$
of the ancillary system are orthogonal, the final density matrix of
the ancillary system can also be represented by a Bloch vector $\boldsymbol{r}=[r_{x},r_{y},r_{z}]$,
which can be derived as
\begin{align}
r_{x} & =\sin(2\alpha)\left[\Gamma_{r}\cos(\phi)-\Gamma_{i}\sin(\phi)\right],\\
r_{y} & =-\sin(2\alpha)\left[\Gamma_{r}\sin(\phi)+\Gamma_{i}\cos(\phi)\right],\\
r_{z} & =\cos(2\alpha),
\end{align}
where $\Gamma_{r}$ and $\Gamma_{i}$ are the real and imaginary components
of the function
\begin{equation}
\Gamma(t)=\bra[\psp]e^{it\Theta_{2}^{\dagger}}e^{-it\Theta_{1}}\ket[\psp].\label{eq:=000020Gamma}
\end{equation}
Here $\ket[\psp]$ denotes the initial state of the probe system and
the operator $\Theta_{k}$ is defined as
\begin{equation}
\Theta_{k}=\wp\hp+\wa h_{k}I+ga_{k}B,\label{eq:=000020defi=000020Gk}
\end{equation}
satisfying the relation
\begin{equation}
H_{\text{tot}}\left(\ket[\psp]\otimes\ket[a_{k}]\right)=\left(\Theta_{k}\ket[\psp]\right)\otimes\ket[a_{k}],
\end{equation}
and since $\ket[a_{k}]$ is the eigenstate of both $\ha$ and $A$,
therefore, we have
\begin{equation}
e^{-itH_{\text{tot}}}\ket[\psp]\otimes\ket[a_{k}]=e^{-it\Theta_{k}}\ket[\psp]\otimes\ket[a_{k}].\label{eq:=000020H-Gk}
\end{equation}

The quantum Fisher information can hence be written in the Bloch representation,
\begin{equation}
F_{Q}=\sin^{2}(2\alpha)\frac{\left(\partial_{\wp}\Gamma_{r}\right)^{2}+\left(\partial_{\wp}\Gamma_{i}\right)^{2}-\left(\Gamma_{r}\partial_{\wp}\Gamma_{r}-\Gamma_{i}\partial_{\wp}\Gamma_{i}\right)^{2}}{1-\Gamma_{r}^{2}-\Gamma_{i}^{2}}.\label{eq:=000020FQ=000020theta}
\end{equation}
where $\partial_{\wp}$ denotes the derivative with respect to the
estimated parameter $\wp$. It is obvious that the optimal $\alpha$
of the initial state of the ancillary system that maximizes the quantum
Fisher information $F_{Q}$ is $\alpha=\pi/4$. Furthermore, as Eq.
\eqref{eq:=000020FQ=000020theta} is independent of $\phi$, we set
the azimuthal angle $\phi$ to be zero and then the optimal initial
state of the ancillary system is given by
\begin{equation}
\ket[\psa]=\frac{1}{\sqrt{2}}\ket[a_{1}]+\frac{1}{\sqrt{2}}\ket[a_{2}].
\end{equation}
By employing the optimal initial state of the ancillary system, the
quantum Fisher information \eqref{eq:=000020FQ=000020theta} can be
simplified as
\begin{equation}
F_{Q}(t)=\left|\partial_{\wp}\Gamma(t)\right|^{2}+\frac{1}{4}\frac{\left[\partial_{\wp}\left|\Gamma(t)\right|^{2}\right]^{2}}{1-\left|\Gamma(t)\right|^{2}}.\label{eq:=000020FQ-Gamma}
\end{equation}

According to Eq. \eqref{eq:=000020FQ-Gamma}, the time scaling of
the quantum Fisher information is determined by $\partial_{\wp}\Gamma(t)$
and $\partial_{\wp}\left|\Gamma(t)\right|^{2}$. Specifically, the
derivative of the function $\Gamma(t)$ with respect to $\wp$ is
related to the derivative of the operator $e^{-it\Theta_{k}}$ with
respect to $\wp$, which can be worked out as
\begin{align}
\partial_{\wp}e^{-it\Theta_{k}}= & -i\int_{0}^{t}e^{-i\left(t-\tau\right)\Theta_{k}}\left(\partial_{\wp}\Theta_{k}\right)e^{-i\tau\Theta_{k}}d\tau\\
= & -ie^{-it\Theta_{k}}\int_{0}^{t}e^{i\tau\Theta_{k}}\hp e^{-i\tau\Theta_{k}}d\tau.
\end{align}
By utilizing the spectral decomposition of $\Theta_{k}$, $\Theta_{k}=\sum_{i}\lambda_{i}^{(k)}\ket[\lambda_{i}^{(k)}]\bra[\lambda_{i}^{(k)}]$,
the integration over the equal eigenvalues of $\Theta_{k}$ may result
in a linear term with the evolution time $t$ while integrating over
$\Theta_{k}$'s unequal eigenvalues yields a constant-order term:
\begin{equation}
\begin{aligned} & \int_{0}^{t}e^{i\tau\Theta_{k}}\hp e^{-i\tau\Theta_{k}}d\tau\\
= & t\sum_{\lambda_{i}^{(k)}=\lambda_{j}^{(k)}}\ket[\lambda_{i}^{(k)}]\bra[\lambda_{i}^{(k)}]\hp\ket[\lambda_{j}^{(k)}]\bra[\lambda_{j}^{(k)}]+O(1).
\end{aligned}
\label{eq:int}
\end{equation}
Consequently, the term $\partial_{\wp}\Gamma(t)$ exhibits linear
scaling behavior with respect to the evolution time $t$, that is
$\partial_{\wp}\Gamma(t)\propto t$, if the linear term in Eq. \eqref{eq:int}
does not vanish. Substituting this result to Eq. \eqref{eq:=000020FQ-Gamma},
we arrive at the Heisenberg scaling of the quantum Fisher information
$F_{Q}$ with respect to time, $F_{Q}(t)\propto t^{2}$.

So far, we have seen the possibility that the quantum Fisher information
achieves the Heisenberg scaling with respect to the evolution time
by employing the optimal initial state of the ancillary system. Nevertheless,
the Heisenberg scaling in terms of the total probe number is more
involved, which is influenced by both the initial state of the probe
system and the interaction term. Details will be investigated in the
following subsection, where the condition for the Heisenberg scaling
with respect to the evolution time will also be elaborated.

\subsection{Configuration of probe system\protect\label{subsec:The-configuration-of}}

In order to present a detailed and quantitative condition for the
Heisenberg scaling with respect to the number of probes, we consider
a system composed of $N$ identical probes, which are uncorrelated
before the unitary evolution under the Hamiltonian. In this case,
the $N$-probe state is $\ket[\psp]=\ket[{\php[(1)]}]\otimes\ket[{\php[(2)]}]\otimes...\otimes\ket[{\php[(N)]}]$,
where $\ket[{\php[(k)]}]$ represents the initial state of the $k$-th
probe. Similarly, the dynamical evolution can be decomposed into one-body
evolutions acting on each probe individually.

Suppose each individual probe of the probe system is initially in
the same state $\ket[{\php}]$ and the Hamiltonian can be written
by the vector of Pauli matrices $\vt[\sigma]$ as
\begin{equation}
\hp=\sum_{i}\vt[m]\cdot\vt[\sigma]^{(i)},\,B=\sum_{i}\vt[n]\cdot\vt[\sigma]^{(i)},
\end{equation}
with $\vt[m]=\left(m_{x},m_{y},m_{z}\right)\in\mathbb{R}^{3}$ and
$\vt[n]=\left(n_{x},n_{y},n_{z}\right)\in\mathbb{R}^{3}$. Note that
we do not specify the free Hamiltonian $\ha$ and the observable $A$
for the ancilla in the total Hamiltonian \eqref{eq:htot}, but require
them to commute with each other only so that they share a common set
of eigenstates. Consequently, the $N$-body operator $\Theta_{k}$
\eqref{eq:=000020defi=000020Gk} can be represented by the Kronecker
sum of one-body operators $\vartheta_{k}^{(i)}$, i.e., $\Theta_{k}=\oplus_{i}\vartheta_{k}^{(i)}$,
where
\begin{equation}
\vartheta_{k}^{(i)}=\lambda\wp\vt[m]\cdot\vt[\sigma]^{(i)}+\frac{\wa}{N}h_{k}I_{2}+ga_{k}\vt[n]\cdot\vt[\sigma]^{(i)}.
\end{equation}
Here $\lambda$ can be considered as the difference between the maximal
and the minimal levels of a single probe, which is currently simulated
by a two-level system. The operator $\vartheta_{k}^{(i)}$ is identical
for each probe, and we denote them as $\vartheta_{k}$ for simplicity
in the following if no ambiguity occurs. Similar to the definition
and relation of $\Theta_{k}$ presented in Eq. \eqref{eq:=000020defi=000020Gk}
and Eq. \eqref{eq:=000020H-Gk}, the one-body operator $\vartheta_{k}$
satisfies the relation
\begin{equation}
e^{-itH_{\text{tot}}}\ket[\psp]\otimes\ket[a_{k}]=\left(e^{-it\vartheta_{k}}\ket[{\php}]\right)^{\otimes N}\otimes\ket[a_{k}].
\end{equation}
We can therefore define the function $\gamma(t)$ for a single probe
by the one-body operator $\vartheta_{k}$ and the initial state of
each probe $\ket[{\php}]$ as
\begin{equation}
\gamma(t)=\bra[{\php}]e^{it\vartheta_{2}}e^{-it\vartheta_{1}}\ket[{\php}],\label{eq:=000020SGamma}
\end{equation}
satisfying
\begin{align}
\Gamma(t) & =\gamma(t)^{N},\label{eq:=000020Gamma-Sgamma}\\
\partial_{\wp}\Gamma(t) & =N\gamma(t)^{N-1}\partial_{\wp}\gamma(t).\label{eq:=000020pGamma-Sgamma}
\end{align}

The quantum Fisher information can then be written in terms of the
function $\gamma(t)$: 
\begin{equation}
F_{Q}(t)=N^{2}\left|\gamma(t)\right|^{2N-2}\left|\partial_{\wp}\gamma(t)\right|^{2},\label{eq:=000020FQ-SGamma}
\end{equation}
which involves a quadratic factor of $N$ that is necessary to the
Heisenberg scaling of the quantum Fisher information with respect
to the total number $N$ of the probes.

However, the absolute value of the function $\gamma(t)$, i.e., $\abs[\gamma(t)]$,
can range between $0$ and $1$ by the definition \eqref{eq:=000020SGamma}
since the dynamical operators $e^{-it\vartheta_{k}}$ are unitary.
If $\left|\gamma(t)\right|\neq1$, the term $\left|\gamma(t)\right|^{2N-2}$
will result in an exponential decay of $F_{Q}(t)$ with respect to
$N$ when $N$ is large, and the quantum Fisher information will be
diminished in this case. So it is critical to guarantee $\left|\gamma(t)\right|=1$
by proper optimizations over the initial states and the Hamiltonians
of the probes and the ancillary system in order to realize the Heisenberg
scaling. This will be investigated in detail in the next subsection.

\subsection{Optimization of time points and probe state}

The quantum Fisher information has a periodic pattern with respect
to the evolution time $t$ originated from the periodic structure
of $\gamma(t)$ \eqref{eq:=000020SGamma} which determines the quantum
Fisher information by Eq. \eqref{eq:=000020FQ-SGamma}. This will
be made clearer later in this subsection. So, we just need to consider
the scaling of the quantum Fisher information with respect to the
specific time points that extremize the quantum Fisher information,
and the measurements on the ancillary system can be performed at those
time points. Furthermore, the initial state of the probe system, denoted
by $\ket[{\php}]$, can also significantly affect the quantum Fisher
information. It will be shown that by optimizing the initial state
of the probe system and selecting appropriate time points for measurements,
it is possible to achieve the Heisenberg scaling of the quantum Fisher
information with respect to time at those time points.

By the Bloch representation of the initial probe state,
\begin{equation}
\ket[{\php}]\bra[{\php}]=\frac{1}{2}\left(I+\vt[v]\cdot\vt[\sigma]\right),\,\vt[v]=\left(v_{x},v_{y},v_{z}\right)\in\mathbb{R}^{3},
\end{equation}
with $|\vt[v]|=1$ for a pure state, the function $\gamma(t)$ can
be written as
\begin{equation}
\gamma(t)=e^{-it\wa\frac{h_{1}-h_{2}}{N}}\left[\gamma_{r}(t)+i\gamma_{i}(t)\right],
\end{equation}
where
\begin{align}
\gamma_{r}(t)= & \cos\left(\mu_{1}t\right)\cos\left(\mu_{2}t\right)\nonumber \\
 & +\frac{\mu_{1}^{2}+\mu_{2}^{2}-g^{2}\left(a_{1}-a_{2}\right)^{2}}{2\mu_{1}\mu_{2}}\sin\left(\mu_{1}t\right)\sin\left(\mu_{2}t\right),\label{eq:gamma-re}\\
\gamma_{i}(t)= & \vt[k](t)\cdot\vt[v],\label{eq:gamma-im}
\end{align}
and
\begin{align}
\vt[k](t)= & c_{m}\vt[m]+c_{n}\vt[n]+c_{mn}\vt[m]\times\vt[n],\label{eq:=000020kt}
\end{align}
with
\begin{align}
\hspace{-0.3cm}c_{m} & =\left[\frac{\cos\left(\mu_{1}t\right)\sin\left(\mu_{2}t\right)}{\mu_{2}}-\frac{\sin\left(\mu_{1}t\right)\cos\left(\mu_{2}t\right)}{\mu_{1}}\right]\lambda\wp,\\
\hspace{-0.3cm}c_{n} & =\left[\frac{a_{2}\sin\left(\mu_{2}t\right)\cos\left(\mu_{1}t\right)}{\mu_{2}}-\frac{a_{1}\sin\left(\mu_{1}t\right)\cos\left(\mu_{2}t\right)}{\mu_{1}}\right]g,\\
\hspace{-0.3cm}c_{mn} & =\frac{\sin\left(\mu_{1}t\right)\sin\left(\mu_{2}t\right)}{\mu_{1}\mu_{2}}\left(a_{1}-a_{2}\right)g\lambda\wp.
\end{align}
Specifically, $\gamma_{r}(t)$ and $\gamma_{i}(t)$ denote the real
and imaginary components of the function $\gamma(t)$, respectively,
up to the global phase $e^{-it\wa\frac{h_{1}-h_{2}}{N}}$ which is
irrelevant to the quantum Fisher information. Here the functions $\mu_{k}$
are defined as
\begin{equation}
\mu_{k}=\sqrt{a_{k}^{2}g^{2}+2a_{k}g\lambda\wp\vt[m]\cdot\vt[n]+\lambda^{2}\wp^{2}},\;k=1,2,
\end{equation}
describing the frequencies of the periodic behavior of the function
$\gamma(t)$, which determines the periodic nature of the evolution
of the quantum Fisher information.

The quantum Fisher information \eqref{eq:=000020FQ-SGamma} indicates
that the exponential decay may substantially counteract the Heisenberg
scaling. However, if $\left|\gamma(t)\right|=1$, the exponential
decay vanishes, allowing for the achievement of the Heisenberg scaling
in terms of the total probe number $N$, so it is crucial to have
$\left|\gamma(t)\right|=1$. In the following, we will explore the
condition to achieve this maximum value of $\left|\gamma(t)\right|$.

The absolute value of the function $\gamma(t)$, which is determined
by the vector $\vt[v]$, can be given by
\begin{align}
\left|\gamma(t)\right|^{2}= & \gamma_{r}^{2}(t)+\gamma_{i}^{2}(t)=\gamma_{r}^{2}(t)+\left(\vt[k](t)\cdot\vt[v]\right)^{2}.\label{eq:absgammat}
\end{align}
In general, when $\gamma_{r}(t)$ and $\vt[k](t)$ are fixed, both
of which are determined by $\mu_{1}$ and $\mu_{2}$, the maximal
$\vt[k](t)\cdot\vt[v]$ over all possible $\vt[v]$ leads to the maximum
value of $\left|\gamma(t)\right|^{2}$. And $\vt[k](t)\cdot\vt[v]$
is maximized when $\vt[v]$ is parallel or antiparallel to $\vt[k](t)$
considering $|\vt[v]|=1$ for the pure initial state $\ket[\psi_{p}]$,
leading to $\vt[k](t)\cdot\vt[v]=\pm\abs[{\vt[k](t)}]$ at the optimal
time points. However, since the initial state of the probe system
does not change over time, this maximum value can be achieved only
at those optimized time points $t_{p}$, and the optimal vector $\vt[v]$
can be obtained at the optimized time points $t_{p}$ by
\begin{equation}
\vt[v]=\frac{\vt[k](t_{p})}{\abs[{\vt[k](t_{p})}]}.
\end{equation}
It can be verified by Eq. \eqref{eq:gamma-re} and \eqref{eq:=000020kt}
that for all time $t$,
\begin{equation}
\gamma_{r}^{2}(t)+\abs[{\vt[k](t)}]^{2}=1.
\end{equation}
So when the time points are optimized, $\left|\gamma(t_{p})\right|=1$,
considering Eq. \eqref{eq:absgammat} and $\vt[k](t_{p})\cdot\vt[v]=\pm\abs[{\vt[k](t_{p})}]$.
This yields the maximum value of $\left|\Gamma(t)\right|^{2}$ at
the optimized time points $t_{p}$: 
\begin{equation}
\left|\Gamma(t_{p})\right|^{2}=\left|\gamma(t_{p})\right|^{2N}=1.
\end{equation}

A further simplification can be achieved by representing the function
$\gamma(t)$ by its amplitude $\abs[\gamma(t)]$ and argument $\theta(t)$,
where
\begin{equation}
\theta(t)=\arctan\frac{\text{Im}\left[\gamma(t)\right]}{\text{Re}\left[\gamma(t)\right]}=\arctan\frac{\gamma_{i}(t)}{\gamma_{r}(t)}.
\end{equation}
As $\left|\gamma(t_{p})\right|=1$ at the optimal time points $t_{p}$,
$\gamma(t_{p})$ can be written as $\gamma(t_{p})=\exp[i\theta(t_{p})]$.
So, the quantum Fisher information at the time points $t_{p}$ can
then be given in a simpler manner: 
\begin{align}
F_{Q}(t_{p})= & N^{2}\abs[\partial_{\wp}\theta(t_{p})]^{2}=N^{2}\left(\frac{\partial_{\wp}\gamma_{i}(t_{p})}{\gamma_{r}(t_{p})}\right)^{2},\label{eq:=000020QFI-theta}
\end{align}
which manifests the Heisenberg scaling in terms of the total probe
number $N$. Moreover, the Heisenberg scaling in terms of the optimal
time points$t_{p}$ can also be derived from Eq. \eqref{eq:=000020QFI-theta}
as the derivative of $\gamma_{i}(t_{p})$ with respect to $\omega$
involves the derivatives of the sine and cosine functions of $\mu_{1}t$
and $\mu_{2}t$ which gives a linear term of $t_{p}$ multiplied by
some coefficient. That coefficient can be further optimized, and the
quantum Fisher information \eqref{eq:=000020QFI-theta} reaches the
maximum when both cosine functions, $\cos\left(\mu_{1}t\right)$ and
$\cos\left(\mu_{2}t\right)$, are equal to $1$ simultaneously. In
this case, we have $\gamma_{i}\left(t_{p}\right)=0$ as $\vt[k](t_{p})=0$,
and thus $\gamma_{r}(t_{p})=1$ as $\left|\gamma(t)\right|=1$ at
the optimal time points. And the quantum Fisher information \eqref{eq:=000020QFI-theta}
can be simplified to
\begin{equation}
F_{Q}(t_{p})=N^{2}\left(\partial_{\wp}\gamma_{i}(t_{p})\right)^{2}.
\end{equation}
Note that
\begin{align}
\partial_{\wp}\theta(t_{p}) & =\partial\gamma_{i}(t_{p})=\partial_{\wp}\vt[k](t_{p})\cdot\vt[v],
\end{align}
so the further optimization of the initial state $\ket[\psi_{p}]$,
i.e., the optimization of the normalized vector $\vt[v]$, can be
given as
\begin{equation}
\vt[v]=\frac{\partial_{\wp}\vt[k](t_{p})}{|\partial_{\wp}\vt[k](t_{p})|},
\end{equation}
and the quantum Fisher information at $t_{p}$ can be obtained as
\begin{align}
F_{Q}(t_{p})= & \frac{\left(a_{1}-a_{2}\right)^{2}t_{p}^{2}N^{2}g^{2}\lambda^{4}\wp^{2}\left[1-\left(\vt[m]\cdot\vt[n]\right)^{2}\right]}{\mu_{1}^{2}\mu_{2}^{2}}.\label{eq:=000020FQtp=000020mn}
\end{align}
Note that this formula only holds at the optimized time points $t_{p}$
where the quantum Fisher information is maximized, implying that Eq.
\eqref{eq:=000020FQtp=000020mn} can be regarded as the upper envelope
of the evolution of quantum Fisher information over time.

It is also noteworthy that as $\gamma_{r}(t)$ is independent of the
initial state $\ket[\psi_{p}]$ of the probe according to Eq. \eqref{eq:gamma-re},
any initial state of the probes can therefore be utilized to achieve
the Heisenberg scaling at those time points unless $\partial_{\wp}\theta(t_{p})=0$
in which case $\gamma(t)$ is independent of the parameter \ensuremath{\omega}
and hence no information about \ensuremath{\omega} can be extracted
from the evolved state of the ancilla according to Eq. \eqref{eq:=000020QFI-theta}.

The above analysis indicates that by selecting appropriate time points
and employing the optimal initial states for both the probe system
and the ancillary system, the maximum quantum Fisher information can
reach the Heisenberg scaling with respect to those time points and
the number of probes, although the probes are initially uncorrelated
as assumed at the beginning of Sec. \ref{subsec:The-configuration-of}.

\subsection{Optimization of probe system}

We consider the Hamiltonians of the ancillary system to be $\ha=A=\sigma_{z}$.
Then the two eigenstates are $\ket[0]$ and $\ket[1]$, and the corresponding
eigenvalues of $A$ are $a_{1}=-a_{2}=1$. Now the quantum Fisher
information at time points $t_{p}$ can be written as
\begin{align}
F_{Q}(t_{p})= & \frac{4t_{p}^{2}N^{2}g^{2}\lambda^{4}\wp^{2}\left[1-\left(\vt[m]\cdot\vt[n]\right)^{2}\right]}{\left(g^{2}+\lambda^{2}\wp^{2}\right)^{2}-\left(2g\lambda\wp\vt[m]\cdot\vt[n]\right)^{2}}.\label{eq:FQtp}
\end{align}
Here only the quadratic term of $t_{p}$ remains.

It can be verified that the maximum of the quantum Fisher information
\eqref{eq:FQtp} can be attained when $\vt[m]\cdot\vt[n]=0$ when
$g$ and $\lambda\wp$ are fixed, implying that noncommuting probe
Hamiltonian $\hp$ and interaction Hamiltonian can increase the estimation
precision when the parameter of the probes is indirectly measured
via the ancillary system alone, which is consistent with the physical
intuition as the interaction Hamiltonian must be noncommuting with
the free Hamiltonian of the probe in order to transfer the information
of the parameter from the probe to the ancillary system. In this case,
the two frequencies coincide,
\begin{equation}
\mu_{1}=\mu_{2}=\sqrt{g^{2}+\lambda^{2}\wp^{2}},
\end{equation}
which leads to the optimal time points as
\begin{equation}
t_{p}=2k\pi/\sqrt{g^{2}+\lambda^{2}\wp^{2}},\,\,k=1,2,3...,
\end{equation}
and the maximum quantum Fisher information is
\begin{equation}
F_{Q}(t_{p})=\frac{4t_{p}^{2}N^{2}g^{2}\lambda^{4}\wp^{2}}{\left(g^{2}+\lambda^{2}\wp^{2}\right)^{2}}.
\end{equation}

If the interaction strength \ensuremath{g} and the coefficient $\lambda\omega$
of the probe Hamiltonian are also allowed to be adjusted, the optimization
of the quantum Fisher information \eqref{eq:FQtp} gives $g=\lambda\wp$
whenever $\vt[m]\cdot\vt[n]\neq\pm1$, not necessarily $\vt[m]\cdot\vt[n]=0$,
which imposes a looser constraint on the probe Hamiltonian and the
interaction Hamiltonian to achieve the highest precision by measurements
on the ancillary system only. The quantum Fisher information at the
optimized time points can be found as
\begin{align}
F_{Q}(t_{p})= & \lambda^{2}N^{2}t_{p}^{2}.\label{eq:=000020FQtp=000020optimal}
\end{align}
If $\vt[m]\cdot\vt[n]=\pm1$, the argument of $\gamma(t)$ becomes
\begin{equation}
\theta(t)=\mp\arctan\left[\tan\left(gt\right)\vt[m]\cdot\vt[v]\right],
\end{equation}
independent of the estimated parameter \ensuremath{\omega}, resulting
in$F_{Q}(t)=0$ for all time $t$, indicating that the Heisenberg
scaling cannot be achieved by commuting probe Hamiltonian and interaction
Hamiltonian, whatever the interaction strength is.

It is worth noting that in order to achieve the Heisenberg limit of
the time scaling of quantum Fisher information, there should always
exist time points $t_{p}$ satisfying $\vt[k](t_{p})\cdot\vt[v]=\pm\abs[{\vt[k](t_{p})}]$
with a growing evolution time. Considering $\vt[k](t_{p})$ is composed
of sine and cosine functions with two frequencies $\mu_{1}$ and $\mu_{2}$while
$\vt[v]$ is independent of time, a straightforward approach to satisfy
this condition is to choose $t_{p}$ as the common period of the two
periodic functions, which can be determined by the least common multiples
of $\mu_{1}^{-1}$ and $\mu_{2}^{-1}$ multiplied by a positive integer,
so that $\vt[k](t_{p})$ coincides with $\vt[v]$ periodically. If
the ratio between the two frequencies $\mu_{1}$ and $\mu_{2}$ is
an integer, the common period is simply the reciprocal of the lower
frequency. If the ratio is a rational number, the least common multiple
of $\mu_{1}^{-1}$ and $\mu_{2}^{-1}$ exists and is equal to the
reciprocal of the greatest common divisor which can be derived by
the Euclidean algorithm. However, if the ratio is an irrational number,
an exact common period does not exist and the two cosine functions
will never synchronize completely in a regular pattern. But one can
always find a rational number to approximate the ratio with an arbitrary
precision, and thus derive an approximate common period for the two
periodic functions to determine the optimal time points $t_{p}$.

\begin{figure}
\subfloat[\centering]{\includegraphics[scale=0.85]{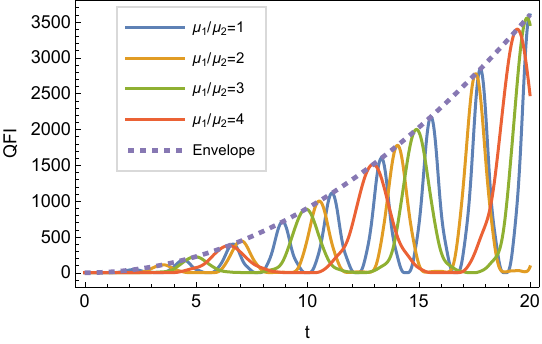}

}

\subfloat[\centering]{\includegraphics[scale=0.88]{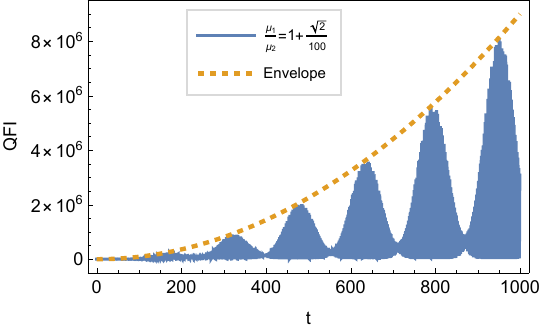}

}

\caption{\protect\label{fig:=000020envelope}The quantum Fisher information
over the evolution time with the optimal initial state of the probes
and the ancillary system and interaction strength $g=\lambda\protect\wp$
for different ratios of two frequencies $\mu_{1}$ and $\mu_{2}$,
compared with the same upper envelope given by Eq. \eqref{eq:=000020FQtp=000020optimal}.
(a) The ratio of two frequencies $\mu_{1}$ and $\mu_{2}$ is rational.
(b) The ratio of the two frequencies is irrational. Parameters: $N=3$,
$\lambda=1$, $\omega=1$, $\protect\wa=2$ and $g=1$.}
\end{figure}

In Fig. \ref{fig:=000020envelope}, the quantum Fisher information
is illustrated with the optimal initial states of the probes and the
ancillary system and the interaction strength $g=\lambda\wp$ for
different ratios between $\mu_{1}$ and $\mu_{2}$, considering both
rational and irrational cases for the ratio, and compare them with
the upper envelope given by Eq. \eqref{eq:=000020FQtp=000020optimal},
showcasing the attainability of the same time scaling for the peaks
of the quantum Fisher information with different $\vt[m]\cdot\vt[n]$
by tuning the interaction strength to $g=\lambda\wp$ and the time
scaling reaches the Heisenberg limit.

Before the conclusion of this section, we remark that this protocol
is distinct from the traditional methods in attaining the Heisenberg
scaling. It can be verified by tracing out the ancilla from the evolved
joint state of the probes and the ancilla that the probes always remain
in a separable mixed state (though not a product state) during the
evolution, while the conventional protocols usually require entangled
states in order for high precision of measurements. In essence, the
key to enhancing measurement precision in this protocol lies in the
interaction. Interaction has been shown useful in enhancing the sensitivity
of quantum measurements \citep{boixo2007generalized,boixo2008quantumlimited,boixo2008quantum,napolitano2011interactionbased,hou2021superheisenberg}.
Nevertheless, those interaction-based quantum metrology protocols
introduce interactions within the probes which are subject to the
parameter-dependent Hamiltonians, while the current protocol couples
the probes to ancillary degrees of freedom which undergo a parameter-independent
evolution. Such a system extension strategy was employed for entanglement-based
quantum metrology protocols before \citep{he2021quantum,demkowicz-dobrzanski2014usingentanglement},
but is new to the interaction-based protocols. By the combination
of the interaction and the system extension strategies, the current
protocol optimizes the measurement sensitivity without entanglement
and expands the possibilities for quantum metrology.

It is also worth mentioning that as this protocol only needs to implement
single-qubit and two-qubit Hamiltonians, it is realizable by the state-of-the-art
quantum technologies on various quantum systems. For example, ion
traps \citep{porras2004effective,lu2019globalentangling,bohnet2016quantum},
superconducting qubits \citep{yan2019strongly,song2019generation,roushan2017spectroscopic,ye2019propagation},
and solid-state spin systems \citep{xie2021beating} have been developed
into platforms that can realize qubit systems and single-qubit and
two-qubit controls with considerable coherence times, operation precisions,
and scalability in recent years. While experimental platforms are
generally vulnerable to environmental noise, technical imperfections,
and control errors, significant advancements have been made in achieving
high fidelities for single-qubit and two-qubit gates across multiqubit
systems. Experiments in solid-state platforms have demonstrated single-qubit
gate fidelities exceeding 99.9\% \citep{yang2019silicon,yoneda2018aquantumdot}
and two-qubit gate fidelities above 99\% \citep{tanttu2024assessment,huang2019fidelity}.
These high fidelities are crucial to the robustness of quantum operations
and various quantum technologies.

\subsection{Effects of uncorrelated noise\protect\label{sec:noise}}

The presence of environmental noise is generally inevitable in practice
and can result in the decoherence of quantum systems, which is detrimental
to the metrological performance of quantum measurements and may lead
the measurement to a lower precision, typically the standard quantum
limit, instead of the Heisenberg scaling. So it is important to study
the impact of noise on the current protocol. For typical quantum systems
that can realize qubits \citep{yan2019strongly,song2019generation},
the main decoherence channels include relaxation and dephasing, etc.
Since the dephasing time is usually shorter than the relaxation time
\citep{houck2008controlling,kakuyanagi2007dephasing,yan2016theflux},
the dephasing channel can have a more significant impact on qubit
systems. So in this subsection, we focus on the dephasing process
as an example to analyze the effects of noise on our protocol.

In the presence of noise channels, the joint dynamical evolution of
the probes and the ancillary system can be described by a master equation:
\begin{equation}
\partial_{t}\rho_{pa}(t)=-i\left[H_{\text{tot}},\rho_{pa}(t)\right]+\sum_{k}\gamma_{k}D\left[L_{k}\right]\left(\rho_{pa}(t)\right),
\end{equation}
where $D\left[L_{k}\right]\left(\cdot\right)=L_{k}\left(\cdot\right)L_{k}^{\dagger}-\frac{1}{2}\left\{ L_{k}^{\dagger}L_{k},\cdot\right\} $
with $L_{k}$the quantum dissipative operators and $\gamma_{k}$ the
noise rates. In the following, we will drop the subscript $k$ from
$\gamma_{k}$ and $L_{k}$for simplicity if there is only one noise
channel under consideration. 

We consider the optimized configurations of the interaction Hamiltonian,
specifically, $\vt[m]\cdot\vt[n]=0$ and $g=\lambda\omega$. The initial
states of the probes and ancillary system are assumed to be the optimal
initial state obtained in the previous section. The reduced dynamics
of the ancillary system can be solved from the above master equation
as
\begin{align}
\rho_{a}(t)= & \frac{1}{2}\left[\begin{array}{cc}
1 & \left(R\right)^{N}\\
\left(R^{*}\right)^{N} & 1
\end{array}\right],\label{eq:rat}
\end{align}
where $R$ is the parameter associated with the unknown frequency
$\omega$. For the sake of simplicity, we consider the case where
the noise is not strong, i.e., the rate $\gamma$ is small and the
number of the probes $N$ is large, and the parameter $R$ can then
be expanded to the first order of $\gamma$ and the zeroth order of
$1/N$ as
\begin{align}
R= & \frac{e^{-\frac{4\gamma t\left(g^{2}+\Omega^{2}\right)}{\Omega^{2}}}}{\Omega^{2}}\left\{ e^{4\gamma t}(\Omega^{2}-g^{2})\right.\nonumber \\
 & \left.+ge^{\frac{6\gamma g^{2}t}{\Omega^{2}}}\left[g\cos(2t\Omega)-i\Omega\sin(2t\Omega)\right]\right\} \label{eq:r}\\
 & +\frac{\gamma ge^{-\frac{4\gamma t\left(g^{2}+\Omega^{2}\right)}{\Omega^{2}}}}{\Omega^{5}}\left\{ -2ie^{4\gamma t}\Omega(\Omega^{2}-g^{2})\right.\nonumber \\
 & +\left.e^{\frac{6\gamma g^{2}t}{\Omega^{2}}}\left[g\left(4\Omega^{2}-3g^{2}\right)\sin(2t\Omega)+2i\Omega(\Omega^{2}-g^{2})\cos(2t\Omega)\right]\right\} ,\nonumber 
\end{align}
where $\Omega=\sqrt{g^{2}+\lambda^{2}\omega^{2}}$.

The quantum Fisher information of the ancillary system with respect
to the frequency \ensuremath{\omega} of the probes can be obtained
from the reduced density matrix \eqref{eq:rat} as
\begin{align}
F_{Q}= & \frac{N^{2}\left(\abs[R]^{2\left(N-2\right)}\right)}{1-\abs[R]^{2N}}\\
 & \times\left[\frac{1}{4}\left(R^{*}\partial_{\wp}R-R\partial_{\wp}R^{*}\right)^{2}\abs[R]^{2N}+\abs[R]^{2}\abs[\partial_{\wp}R]^{2}\right].\nonumber 
\end{align}
Although the quantum Fisher information is oscillating with time,
it can be observed that there is only a single frequency, so one can
obtain the upper envelope of the quantum Fisher information:
\begin{align}
F_{\text{env}}\simeq & \lambda^{2}N^{2}t^{2}\left[4^{1-N}e^{-6\gamma Nt}\left(e^{\gamma t}+1\right)^{2N-2}\right].\label{eq:=000020Fev}
\end{align}
Here only the lowest quadratic term of $t$ remains.

It can be observed that the upper envelope of the quantum Fisher information
increases with time initially and then decreases exponentially after
a while according to Eq. \eqref{eq:=000020Fev}. Thus there exists
an optimal time point $t_{\text{op}}$ where the quantum Fisher information
reaches the maximum. It can be derived by maximizing $F_{\text{env}}$
over the time $t$ that the optimal time point $t_{\text{op}}$ is
\begin{equation}
t_{\text{op}}=\frac{2}{5N\gamma}+O(\frac{1}{N^{2}}),\label{eq:top-1}
\end{equation}
which can be understood as the effective time for the Heisenberg scaling,
as the quantum Fisher information falls exponentially beyond this
time point. This will be further verified later. The maximum of the
quantum Fisher information envelope at $t_{\text{op}}$ is then given
by
\begin{align}
F_{\text{env}}(t_{\text{op}})\simeq & \frac{4\lambda^{2}e^{-\frac{9}{25N}-2}}{25\gamma^{2}},\label{eq:feop}
\end{align}
which is independent of $N$ when $N$ is large. Note that when $\gamma\rightarrow0$,
the quantum Fisher information \eqref{eq:feop} goes to infinity.
This is because when $\gamma\rightarrow0$, it becomes the noiseless
case and the quantum Fisher information can always increase with time,
so the optimal time point does not exist in this case, i.e., the optimal
time $t_{\text{op}}$ is $+\infty$, and the quantum Fisher information
becomes infinitely large when it is maximized over the evolution time.

In order to reach the maximum quantum Fisher information, the condition
\begin{equation}
t_{\text{op}}>\tau/2\label{eq:top}
\end{equation}
needs to be satisfied, where $\tau=\pi/\Omega$ is the period of the
quantum Fisher information due to the oscillation of $R$ in Eq. \eqref{eq:r}.
For a small $\gamma$, the solution to the inequality \eqref{eq:top}
is 
\begin{equation}
\gamma<\frac{4\sqrt{2}\lambda\omega}{5\pi N},
\end{equation}
indicating how small the decay rate $\gamma$ needs to be in order
for the maximum quantum Fisher information \eqref{eq:feop} to be
attainable.

\begin{figure}
\includegraphics[scale=0.9]{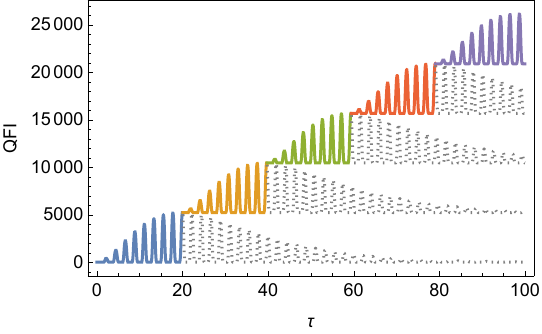}

\caption{\protect\label{fig:mqfi}The quantum Fisher information for a total
interrogation time $T\protect\geq t_{\text{op}}$ in the presence
of dephasing noise. In this case, the maximum of the quantum Fisher
information increases linearly with $N$ and $T$ approximately. The
total interrogation time $T$ is divided into multiple segments, each
a duration of the optimal time $t_{\text{op}}$ for a single segment,
and the evolution is restarted with fresh probes and ancilla at the
beginning of each segment. The gray dotted lines indicate that the
evolutions beyond the optimal time points are replaced by new evolutions
with fresh probes and ancilla when the maximum of quantum Fisher information
is reached in a single segment, and the true evolutions are represented
by the colored solid lines. Parameters: $N=10$, $\gamma=10^{-3}$
and $g=\lambda\omega=1$.}
\end{figure}

As the quantum Fisher information turns to drop with time beyond the
optimal time point $t_{\text{op}}$, the best strategy to increase
the measurement precision for a long interrogation time $T$, $T>t_{\text{op}}$,
is to divide $T$ into multiple segments, each a duration of $t_{\text{op}}$,
and the evolution is restarted with fresh probes and ancilla at the
beginning of each segment. The number of segments is $m=T/t_{\text{op}}$,
and the maximum quantum Fisher information for the total interrogation
time $T$ is given by
\begin{align}
F_{\text{env}}(T)=mF_{\text{env}}(t_{\text{op}})\simeq & \frac{2\lambda^{2}NT}{5e^{2}\gamma}
\end{align}
due to the additivity of the quantum Fisher information for the independent
segments of the evolution. This result is illustrated in Fig. \ref{fig:mqfi},
exhibiting an approximate linear scaling of the quantum Fisher information
with respect to $T$ and suggesting the uncorrelated noise decreases
the measurement precision to the standard quantum limit in this case.

In contrast, if the total interrogation time $T$ is shorter than
$t_{\text{op}}$, we have
\begin{align}
F_{\text{env}}(T)= & \lambda^{2}N^{2}T^{2}\left[4^{1-N}e^{-6\gamma NT}\left(e^{\gamma T}+1\right)^{2N-2}\right].
\end{align}
As the exponentials $e^{-6\gamma NT}$ and $e^{\gamma T}$ keep almost
constant according to Eq. \eqref{eq:top-1} and $e^{\gamma T}\rightarrow1$
when $N$ is large, the quantum Fisher information can approximately
maintain the Heisenberg scaling in terms of $T$ and $N$ within this
regime. This indicates that $t_{op}$ characterizes the effective
time for the Heisenberg scaling of the current protocol.

\section{Conclusions\protect\label{sec:Conclusion}}

In this paper, we propose a protocol that allows uncorrelated probes
to surpass the standard quantum limit in parameter estimation by introducing
parameter-independent interaction of the probes to external degrees
of freedom which serve as an ancillary system. The measurement procedure
is simplified by focusing on the ancillary system alone, which is
not necessarily the optimal strategy but turns out to be sufficient
to achieve the Heisenberg scaling in the precision. The quantum Fisher
information exhibits periodic behavior with the evolution time, suggesting
that finding appropriate time points is critical to the maximization
of the quantum Fisher information. To achieve the highest precision,
we perform optimizations of the quantum Fisher information by tailoring
the interaction Hamiltonian, finding proper time points for measurements,
and tuning the interaction strength, and finally arrive at the Heisenberg
scaling for the estimation precision of the parameter in the probe
Hamiltonian in terms of both the evolution time and the total probe
number. Moreover, we consider the impact of noise on the quantum Fisher
information, and analyze the dephasing process as an example to show
the effect of noise on the precision of measurements.

As a contrast, for a standard measurement protocol to achieve the
Heisenberg scaling in the parameter estimation, it typically prepares
the probes in an entangled or squeezed state, followed by an encoding
unitary evolution process and subsequent measurements on the final
state. The outcomes of these measurements are then analyzed to estimate
the parameters of interest. However, the utilization of quantum resources
is usually challenging in practical applications. The advantage of
our protocol is that the Heisenberg scaling can be achieved by a product
state of the probes and only the ancilla needs to be measured, which
simplifies both the preparation of the initial state and the implementation
of the measurement strategy. This simplicity of the protocol may reduce
the complexity of realizing Heisenberg-limited precision in experiments
and real applications in quantum metrology and other fields.
\begin{acknowledgments}
This work is supported by National Natural Science Foundation of China
(Grant No. 12075323) and the Innovation Program for Quantum Science
and Technology (Grant No. 2021ZD0300702).
\end{acknowledgments}

\bibliographystyle{apsrev4-2}
\bibliography{Heisenberglimit}

\end{document}